%% file: paper.tex
\documentclass[12pt,a4paper]{article}
\usepackage[hmargin=21mm,vmargin=21mm]{geometry}
\usepackage[T1]{fontenc}
\usepackage[utf8]{inputenc}
\usepackage{graphicx}
\graphicspath{{./plots/}}
\usepackage{natbib}

\usepackage{amsmath}
\usepackage{amsfonts,amssymb}
\usepackage{mathtools}
\usepackage{bm}

\usepackage[sc]{mathpazo}
\usepackage[scaled=.95]{helvet}
\usepackage{courier}
\usepackage{url}

\usepackage{setspace} 
\usepackage[nolists]{endfloat}

\DeclareMathOperator{\logit}{logit}
\DeclareMathOperator{\E}{E}
\DeclareMathOperator{\var}{var}
\newcommand{\Fpen}{\mathbf{F}_{\text{pen}}}
\newcommand{\indy}{\mathbf{1}_{\{0\}}(y_{rt})}

\title{A zero-inflated endemic-epidemic model with an application to
  measles time series in Germany}
\author{Junyi Lu and Sebastian Meyer\thanks{
    Correspondence: Sebastian Meyer,
    Institute of Medical Informatics, Biometry, and Epidemiology,
    Waldstr. 6, 91054 Erlangen, Germany.
    E-mail: \texttt{seb.meyer@fau.de}}}
\date{\itshape\small
Institute of Medical Informatics, Biometry, and Epidemiology,\\
Friedrich-Alexander-Universit\"at Erlangen-N\"urnberg, Erlangen, Germany
}

\begin{document}

\maketitle

\begin{abstract}\noindent
Count data with excessive zeros are often encountered when modelling
infectious disease occurrence. The degree of zero inflation can vary over
time due to non-epidemic periods as well as by age group or region.
The existing endemic-epidemic modelling framework (\emph{aka} HHH)
lacks a proper treatment for surveillance data with excessive zeros as it
is limited to Poisson and negative binomial distributions.
In this paper, we propose a multivariate zero-inflated endemic-epidemic
model with random effects to extend HHH.
Parameters of the new zero-inflation and the HHH part of the model can be
estimated jointly and efficiently via (penalized) maximum likelihood
inference using analytical derivatives.
A simulation study confirms proper convergence and
coverage probabilities of confidence intervals.
Applying the model to measles counts in the 16 German states, 2005--2018,
shows that the added zero-inflation improves probabilistic
forecasts.

\end{abstract}

\newpage

\section{Introduction}
Infectious disease models help to understand the mechanisms of disease
spread and can be used to generate forecasts. 
The endemic-epidemic modelling approach of \citet{held.etal2005},
\textit{aka} the HHH model (after the authors' initials),
is frequently adopted for time series of infectious disease counts.
It is motivated by a branching process with immigration in that it
decomposes disease incidence into endemic and autoregressive parts.
The comprehensive R package \texttt{surveillance}
\citep{meyer.etal2017} provides tools for model estimation, simulation and visualization.
The HHH model has been applied to a variety of infectious diseases, including
norovirus gastroenteritis \citep{meyer.held2017},
invasive pneumococcal disease \citep{chiavenna.etal2019},
pertussis \citep{munro.etal2020},
and COVID-19 \citep{dickson.etal2020,Giuliani2020,ssentongo.etal2020}.

Count data with excessive zeros are often encountered in public health surveillance
of rare diseases, 
such as syphilis in the USA \citep{Yang2013}, 
visceral leishmaniasis at block level in India \citep{Nightingale2020},
or dengue fever in China \citep{Wang2014} or Brazil \citep{Schmidt2011}.
Zero-inflated and hurdle models are typically used
to analyze such data.
\citet{Rose2006} conclude that the choice between a zero-inflated 
and a hurdle model should generally be driven by the study purpose
and the assumed underlying process.
A zero-inflated model, as a mixture model,
assumes two underlying disease processes: 
zeros are generated both from the at-risk population (sampling zeros) and
the not-at-risk population (structural zeros).
A hurdle model, on the other hand, 
considers all zeros coming from the at-risk population. 
For counts of infections,
zero-inflated models are more appropriate than hurdle models
because a large not-at-risk population may exist,
e.g., those immunized by vaccination or recovery, 
or a community which is detached from an outbreak area.

To account for the not-at-risk population,
we propose an extension of the HHH approach for infectious disease time series with excessive zeros: 
a multivariate zero-inflated HHH model. 
The zero-inflation part can capture seasonality but also
autoregressive effects to reflect that zeros become more likely
with fewer past counts.
The HHH part of the mixture uses the same endemic-epidemic decomposition
as in the original HHH model.
Region-specific random effects are allowed in both parts
to account for heterogenous reporting or varying infection risks,
for example due to demographic factors not covered by covariates.

This paper is organized as follows. 
Section 2 describes the proposed modelling and inference approach,
which is evaluated with a simulation study in Section 3.
In Section 4 we investigate different zero-inflated models for measles data from Germany
and compare their forecast performance with classical HHH models.
Section 5 concludes the paper.

\section{Model formulation and inference}

\subsection{Endemic-epidemic modelling}

The so-called HHH model
\citep{held.etal2005, paul.held2011} assumes that the number of new
cases $Y_{rt}$ of a (notifiable) infectious disease
in unit $r$, $r = 1, \dots, R$ at time $t$, $t = 1, \dots, T$,
given the past counts,
follows a negative binomial distribution,
\begin{equation*}
Y_{rt}|\mathcal{F}_{t-1} \sim \operatorname{NB}(\mu_{rt},\psi_r),
\end{equation*}
with conditional mean $\mu_{rt}$ and conditional variance $\mu_{rt} (1 + \psi_r \mu_{rt})$.
Here, $\mathcal{F}_{t-1} = \sigma(Y_1,\dots,Y_{t-1})$ represents the
information up to time $t - 1$, and $\psi_r > 0$ are unit-specific
overdispersion parameters, possibly assuming $\psi_r \equiv \psi$.
Units typically correspond to geographical regions, age groups, or
the interaction of both \citep{meyer.held2017}.

The endemic-epidemic modelling approach then decomposes the
infection risk additively into
an autoregressive
component (for within-unit transmission),
a spatiotemporal or neighbourhood component (for transmission from other units),
and an endemic component (for cases not directly linked to previously
observed cases). The endemic component is sometimes called background
risk, environmental reservoir, or immigration component.
More specifically, the expected number of new cases $\mu_{rt}$ is modelled as
\begin{equation} 
\mu_{rt} = \lambda_{rt} y_{r, t-1}+ \phi_{rt} \sum_{q \neq r}
 w_{qr} y_{q, t- 1} + \nu_{rt} ,
\end{equation}
where $\lambda_{rt} > 0$ is the autoregressive parameter,
$\phi_{rt} > 0$ is the spatiotemporal parameter
linking region $r$ to neighbouring regions via transmission weights
$w_{qr}$ (possibly unknown as in \citealp{meyer.held2014}),
and $\nu_{rt} > 0$ is the endemic component.
All three parameters are modelled on the log scale as
\begin{align*}
\log(\lambda_{rt}) &= \alpha^{(\lambda)} + b_r^{(\lambda)} + 
\bm{x}_{rt}^{(\lambda)T} \bm{\beta}^{(\lambda)} + \log\big(o_{rt}^{(\lambda)}\big),\\
\log(\phi_{rt}) &= \alpha^{(\phi)} + b_r^{(\phi)} + 
\bm{x}_{rt}^{(\phi)T} \bm{\beta}^{(\phi)} + \log\big(o_{rt}^{(\phi)}\big),\\
\log(\nu_{rt}) &= \alpha^{(\nu)} + b_r^{(\nu)} + 
\bm{x}_{rt}^{(\nu)T} \bm{\beta}^{(\nu)} + \log\big(o_{rt}^{(\nu)}\big),
\end{align*}
where in each component, $\alpha^{(\cdot)}$ and $b_r^{(\cdot)}$ are fixed
and zero-mean random intercepts, respectively,
$\bm{\beta}^{(\cdot)}$ is a vector of unknown coefficients for the
covariates $\bm{x}_{rt}^{(\cdot)}$,
and $o_{rt}^{(\cdot)}$ is an optional offset,
e.g., population fractions $o_{rt}^{(\nu)}=n_{rt}/n_{.t}$ in the endemic component.

Likelihood inference for HHH models and many of its extensions
is implemented in the R package \texttt{surveillance}
\citep{meyer.etal2017}, which also contains several example datasets and
corresponding vignettes for illustration.

\subsection{Zero-inflated HHH model}

To account for excess zeros in surveillance time series, we propose to
extend the above HHH model to a zero-inflated model, which we will call HHH4ZI.
For this purpose, we assume the number of new cases $Y_{rt}$ given
$\mathcal{F}_{t-1}$ to follow a zero-inflated (ZI) negative binomial
distribution \citep{Yau2003}.
Its probability mass function is given by
\begin{equation}
f_{ZI}(y_{rt}; \mu_{rt}, \psi_r, \gamma_{rt}) =
\gamma_{rt} \cdot \indy + (1 - \gamma_{rt}) \cdot f(y_{rt}; \mu_{rt}, \psi_r),
\end{equation}
which represents a mixture of a point mass at zero and a HHH model with
probability mass function $f(y_{rt}; \mu_{rt}, \psi_r)$, i.e.,
a $\operatorname{NB}(\mu_{rt},\psi_r)$ distribution.
The zero-inflation parameter $\gamma_{rt}$ describes the probability
that a zero count comes from the degenerate distribution.
If $\gamma_{rt} \equiv 0$, 
the mixture model would reduce to a HHH model.
Otherwise, a proportion $\gamma_{rt} \in (0,1)$ of the population is assumed to be
not at risk of infection, while infections in the remaining population
follow a HHH model.

Simple zero-inflated models assume $\gamma_{rt} \equiv \gamma$ with a
single parameter. However, we will typically model the logit-proportion
with a linear predictor,
\begin{equation}
\label{eq:gamma}
\logit(\gamma_{rt}) = \alpha^{(\gamma)} + b_r^{(\gamma)} + 
\bm{x}_{rt}^{(\gamma)T} \bm{\beta}^{(\gamma)},
\end{equation}
including an intercept $\alpha^{(\gamma)}$, random unit-specific deviations
$b_r^{(\gamma)}$, and covariate effects,
similar to the parameters $\lambda_{rt}$, $\phi_{rt}$, and $\nu_{rt}$ of
the HHH mean $\mu_{rt}$.
For example, yearly seasonality can be modelled via
$\bm{x}_{rt}^{(\gamma)} = (\sin(\omega t), \cos(\omega t), \dots, \sin(S \cdot \omega t), \cos(S \cdot \omega t))^\top$,
where $S$ denotes the number of harmonics and $\omega = 2\pi/52$ for weekly data,
or $\omega = 2\pi/26$ for bi-weekly data. 
Note that, for any $\delta,\zeta \in \mathbb{R}$,
\begin{equation}\label{eq:sincos}
\delta \sin(\omega t) + \zeta \cos(\omega t) = A \sin(\omega t + \varphi),
\end{equation}
where $A = \sqrt{\delta^2 + \zeta^2}$ is the amplitude and
$\varphi = \arctan(\zeta / \delta)$ is the phase shift of a sinusoidal wave.
Furthermore, we can relate the zero-inflation probability to past counts
to model that (temporary) low incidence tends to inflate the chance of
zero cases. Suppose $\bm{x}_{rt}^{(\gamma)} = y_{r,t-1}$,
then Equation~\eqref{eq:gamma} can be rewritten in terms of the odds
$
\gamma_{rt}/(1-\gamma_{rt}) = \exp(\alpha^{(\gamma)}+ b_r^{(\gamma)})
\exp(\beta^{(\gamma)} y_{r, t-1})
$
that the count inherits from the degenerate distribution.
One additional case in $y_{r,t-1}$ would change the odds for excess zeros
by a factor of $e^{\beta^{(\gamma)}}$, where we would expect $\beta^{(\gamma)} < 0$.
Note that the odds for excess zeros are intrinsically driven by two opposite forces:
Firstly, with higher incidence in a region the probability of contact
with an outbreak member increases,
such that a larger part of the population is at risk.
Secondly, the population immunized by recovery will increase in the long run.

The conditional mean and variance of the zero-inflated model
can be easily derived from a hierarchical formulation using a latent
Bernoulli variable.
Keeping notation simple by omitting the explicit conditioning on
$\mathcal{F}_{t-1}$ and random effects, we can write
\begin{align*}
W_{rt} &\sim \operatorname{Bernoulli} (\gamma_{rt}), \\
Y_{rt}|W_{rt} &\sim \operatorname{NB} ((1 - W_{rt})\,\mu_{rt},\psi_r).
\end{align*}
By the laws of total expectation and variance,
\begin{equation}
\E(Y_{rt}) = \E_W(E(Y_{rt}|W_{rt}))
= \E_W((1 - W_{rt})\,\mu_{rt})
= (1 - \gamma_{rt})\,\mu_{rt}
\end{equation}
and
\begin{align}
\var (Y_{rt}) 
&= \E_W(\var(Y_{rt}|W_{rt})) + \var_W(\E(Y_{rt}|W_{rt})) \notag\\
&= (1 - \gamma_{rt})\,(1 + \mu_{rt} \psi_r + \gamma_{rt}\mu_{rt})\,\mu_{rt}.
\end{align}

The HHH model allows for the estimation of an effective reproduction
number \citep{Bauer2018}. For this purpose,
the expected non-environmental risk can be written in matrix form as
$\bm{A}_t \bm{y}_{t - 1}$,
where $\bm{y}_t = (y_{1t}, \dots, y_{Rt})^T$
and $\bm{A}_t$ is a $R \times R$ matrix with diagonal elements
 $(\bm{A}_t)_{r,r} = (1 - \gamma_{rt})\lambda_{rt}$
 and $(\bm{A}_t)_{r,r'} = (1 - \gamma_{rt}) \phi_{rt} w_{r' r}$ for $r\ne r'$.
The model-based effective reproduction number $R_t$ is the dominant
eigenvalue of $\bm{A}_t$ \citep[Part II]{Diekmann2012}.

We follow \citet{paul.held2011} by considering two variants for the
distribution of the random effects:
uncorrelated random effects or within-unit correlation between components.
We denote the vector of all random effects from the four components by
$\bm{b} = (\bm{b}^{(\lambda)T}, \bm{b}^{(\phi)T}, \bm{b}^{(\nu)T}, \bm{b}^{(\gamma)T})^T$,
which is assumed to be multivariate normal with mean $\bm{0}$ and covariance
matrix $\bm{\Sigma}$.
The uncorrelated variant is given by
\begin{equation}
\bm{\Sigma} = 
\text{blockdiag}(\sigma_\lambda^2 \mathbf{I}_R,
 \sigma_\phi^2 \mathbf{I}_R, 
 \sigma_\nu^2 \mathbf{I}_R,
  \sigma_\gamma^2 \mathbf{I}_R),
\end{equation}
where $\sigma_\lambda^2, \sigma_\phi^2, \sigma_\nu^2, \sigma_\gamma^2$ 
are unknown variance parameters for each component and $\mathbf{I}_R$
denotes the identity matrix of size $R$.
For correlation between different components,
the covariance matrix is alternatively defined as
\begin{equation}
\bm{\Sigma} = \bm{\Omega} \otimes \mathbf{I}_R,
\end{equation}
where $\bm{\Omega} \in \mathbb{R}^{4 \times 4}$ is an unknown covariance matrix 
and $\otimes$ denotes the Kronecker product. 
The positive definiteness of $\bm{\Sigma}$ is ensured if $\bm{\Omega}$ is positive definite \citep{Horn1991}.
The random effects are still uncorrelated between different units.
In order to ensure computational efficiency and enforce the positive definiteness of $\bm{\Omega}$, 
we use a spherical parametrization \citep{Pinheiro1996, Rapisarda2007}.
Details can be found in the Appendix.

\subsection{Inference}

The log-likelihood of a HHH4ZI model \emph{without} random effects is given by
\begin{equation}
l(\bm{\theta}, \bm{\tilde\psi}) = \sum_{r,t} \log f_{ZI}(y_{rt}; \mu_{rt}, \psi_r, \gamma_{rt}),
\end{equation}
where $\bm{\theta} = (\alpha^{(\lambda)}, \alpha^{(\phi)}, \alpha^{(\nu)}, \alpha^{(\gamma)}, 
\bm{\beta}^{(\lambda)T}, \bm{\beta}^{(\phi)T}, \bm{\beta}^{(\nu)T}, \bm{\beta}^{(\gamma)T})^T$
is the vector of parameters affecting the mean $(1-\gamma_{rt})\mu_{rt}$, and
$\bm{\tilde\psi}$ is the vector of transformed (unit-specific) overdispersion parameters,
using $\tilde{\psi}_r = -\log(\psi_r)$ to allow for
unconstrained optimization.
The parameters can be estimated by numerical maximization of the log-likelihood
$l(\bm{\theta}, \bm{\tilde\psi})$. We use the quasi-Newton algorithm available
as \texttt{nlminb()} in R \citep{R:base}, in conjunction with the analytical
gradient and Hessian.

When fitting a HHH4ZI model \emph{with} random effects, we follow a penalized likelihood approach \citep{Kneib2007}
along the lines of the original HHH model \citep{paul.held2011}.
The penalized log-likelihood is given by
\begin{equation}
\label{eq:lpen}
l_{\text{pen}}(\bm{\theta}, \bm{b}, \bm{\tilde\psi}; \bm{\Sigma}) = l(\bm{\theta}, \bm{b}, \bm{\tilde\psi}) + \log p(\bm{b}|\bm{\Sigma}),
\end{equation}
with
\begin{equation*}
l(\bm{\theta}, \bm{b}, \bm{\tilde\psi}) = \sum_{r,t} \log f_{ZI}(y_{rt}; \mu_{rt}, \psi_r, \gamma_{rt}).
\end{equation*}
As the vector of random effects $\bm{b}$ is assumed to follow a
multivariate normal distribution,
\begin{equation*}
\log p(\bm{b}|\bm{\Sigma}) = -\frac{1}{2}\bm{b}^T\bm{\Sigma}(\bm{\theta})^{-1}\bm{b} + \text{const},
\end{equation*}
where we omit additive terms not depending on $\bm{b}$ when optimizing~\eqref{eq:lpen}.
The penalized score function $\mathbf{s}_{\text{pen}}$
and Fisher information matrix $\Fpen$
used for numerical optimization are given in
the Appendix.

The full marginal likelihood of the variance parameters is given by
\begin{align*}
L_{\text{marg}}(\bm{\Sigma}) &=  \int \exp\{l_{\text{pen}}(\bm{\theta}, \bm{b}, \bm{\tilde\psi}; \bm{\Sigma}) \}
d\bm{\theta} d\bm{\tilde{\psi}} d\bm{b}.
\end{align*}
We apply a Laplace approximation and obtain the marginal log-likelihood
\begin{equation*}
l_{\text{marg}}(\bm{\Sigma}) \approx l(\bm{\hat{\theta}}, \bm{\hat{b}}, \bm{\hat{\tilde\psi}})
-\frac{1}{2} \log |\bm{\Sigma}| 
- \frac{1}{2} \bm{\hat{b}}^T \bm{\Sigma}^{-1} \bm{\hat{b}} 
- \frac{1}{2} \log |\Fpen(\bm{\hat{\theta}},\bm{\hat{b}},\bm{\hat{\tilde{\psi}_i}}; \bm{\Sigma})|,
\end{equation*}
where
$\bm{\hat{\theta}}$, $\bm{\hat{b}}$ and $\bm{\hat{\tilde\psi}}$ 
are estimates based on a given $\bm{\Sigma}$.
As these estimates and thus $l(\bm{\hat{\theta}}, \bm{\hat{b}},
\bm{\hat{\tilde\psi}})$ change only slowly as a function of $\bm{\Sigma}$
\citep{Breslow1993, Kneib2007}, the marginal log-likelihood can be
approximated as
\begin{equation}
\label{eq:lmar}
l_{\text{marg}}(\bm{\Sigma}) \approx 
-\frac{1}{2} \log |\bm{\Sigma}| 
- \frac{1}{2} \bm{b}^T \bm{\Sigma}^{-1} \bm{b} 
- \frac{1}{2} \log |\Fpen(\bm{\theta},\bm{b},\bm{\tilde{\psi}}; \bm{\Sigma})|,
\end{equation}
where $\bm{b}$, $\bm{\theta}$ and $\bm{\tilde{\psi}}$ are current
estimates and do not directly depend on $\bm{\Sigma}$.
We use numerically robust Nelder-Mead optimization as implemented in R's
\texttt{optim()} to maximize~\eqref{eq:lmar}.

Overall, a zero-inflated HHH model with random effects can be estimated
via the following algorithm:
\begin{enumerate}
\item Initialize the covariance matrix $\bm{\Sigma}$ of the random effects.
\item Given $\bm{\Sigma}$, update $\bm{\theta}$, $\bm{b}$,
$\bm{\tilde\psi}$ by maximizing the penalized log-likelihood
(Equation~\ref{eq:lpen}).
\item Given current $(\bm{\theta}, \bm{b}, \bm{\tilde\psi})$,
update the covariance matrix $\bm{\Sigma}$ by maximizing the marginal log-likelihood
(Equation~\ref{eq:lmar}) with respect to its spherical parameters (see Appendix).
\item Iterate steps 2 and 3 until convergence.
\end{enumerate}
An R package implementing this method is provided in the supplementary material.

\section{Simulation}

We conducted a simulation study with $N = 1000$ repetitions for different
time series lengths $T \in \{50,100,500\}$.
The data-generating process was a multivariate HHH4ZI model for the 16
German states using
$\alpha^{(\lambda)} = -0.3$, 
 $\alpha^{(\phi)} = 0.5$, 
$\alpha^{(\nu)} = 0.5$
$\alpha^{(\gamma)} = 0.2$,
zero-inflation terms
$\bm{x}_{rt}^{(\gamma)} = (\sin(2\pi t/26), \cos(2\pi t/26), y_{r,t-1})^T$
with coefficients $\bm{\beta}^{(\gamma)} = (0.4, -0.3, -0.1)^T$,
homogeneous overdispersion $\psi_r = 0.5$, and no random effects.
We assumed normalized first-order transmission weights, i.e.,
$w_{qr}=1/m_q$, if state $q$ (with $m_q$ neighbours) is
adjacent to state $r$, and $w_{qr}=0$ otherwise.
Furthermore, we used a state-specific offset $o_r^{(\phi)}=n_r/n_.$
such that the rate of imported infections
scales with the population size $n_r$. 

Maximum likelihood estimation converged for all simulated datasets.
The estimated parameter vectors are summarized by means and standard
deviations in Table~\ref{tab:sim}.
As expected, increasing the length of the time series improves the
estimates by reducing the variance.
Note that the spatio-temporal parameter $\alpha^{(\phi)}$ cannot be
estimated reliably from short time series ($T = 50$ or 100).
A possible reason is that the spatiotemporal component has a relatively
small impact on the time series in the assumed model.
For long time series ($T = 500$), all parameters are estimated reliably.
The coverage probability of Wald confidence intervals is
approximately equal to their nominal level (95\% or 50\%) for all
parameters and time series lengths.

\input{./tables/sim}

\section{Application}

\subsection{Data}

\begin{figure}[p]
\includegraphics[width=\textwidth]{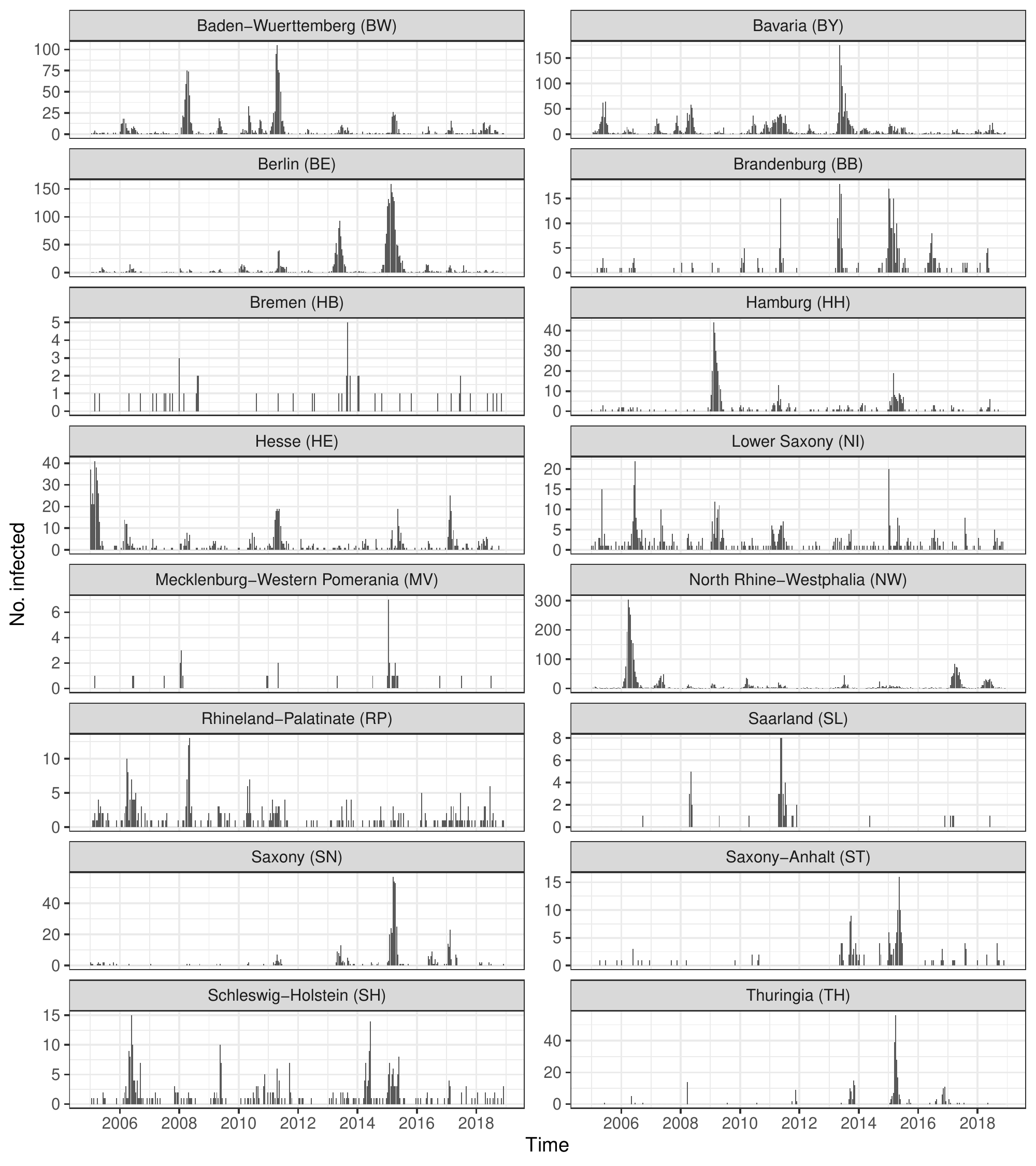}
\centering
\caption{Bi-weekly number of measles cases in the 16 federal states of Germany for the years 2005 to 2018. Note the varying y-axes.
}
\label{fig:measles_ts}
\end{figure}

We apply the proposed HHH4ZI model to count time series
of reported measles cases in the 16 federal states of Germany, from 2005 to 2018.
\citet{herzog.etal2011} have used an earlier version of these data to study the association
 between measles incidence and vaccination coverage.
We follow their approach of aggregating the counts over bi-weekly intervals
to approximately match the generation time of measles of around 10 days
\citep{Fine1982}.
Figure~\ref{fig:measles_ts} shows that measles counts
generally remained at low levels with many zeros during off-seasons. 
However, outbreaks with over 50 cases occurred in several states.
The time series of some states, such as Berlin and Brandenburg,
suggest a weak biennial cycle.
A possible explanation is that susceptibles in low-immune (e.g.,
anthroposophic) communities \citep{Ernst2011} are depleted during epidemic years.
Only minor epidemics may occur throughout the following year when the
susceptible population is replenished by births \citep{Dalziel2016}.

Vaccination coverage is surveyed by local health authorities by 
checking the receipt of the first and second doses of measles-mumps-rubella (MMR) vaccination 
among school starters. 
We assume the coverage rate among children who don't present a vaccination card at
the day of the medical examination (6--14\% on average for the different
states) to be
half of that of children presenting a vaccination card \citep{herzog.etal2011}.
These data are currently available until 2018 (before measles
vaccination became mandatory in Germany), which is why we restrict our
measles analysis to the period from 2005 to 2018.

\input{./tables/fact}

In Table~\ref{tab:fact} we compare summary statistics of case counts and
vaccination coverage among federal states. 
The maximum bi-weekly counts range from only 5 (Bremen) 
to 304 (North Rhine-Westphalia). 
Periods without any measles cases are rarest in Bavaria (66 bi-weeks), 
whereas Mecklenburg-Western Pomerania observed no cases at 340 (93.4\%) of
the 364 time points.
Total case numbers correlate strongly with population size, but Berlin
experienced relatively large epidemics.
The yearly updated state population will serve as an offset in the endemic
model component. Relative changes from the beginning to the end of the
study period are in the range from -11\% (Saxony-Anhalt) to +7\% (Berlin).
The estimated vaccination coverage varies from
86\% (Bremen, 2005) to above 95\% (Thuringia, 2008--2012).
Germany's five eastern states (BB, MV, SN, ST, TH) already
had a high initial vaccination coverage for historical reasons (measles
vaccination was mandatory in the former German Democratic Republic).
In the other states, vaccination coverage tended to increase during the
investigated period (see Figure~S1 in the supplementary material).

\subsection{Models}

We first consider the simple Poisson HHH model, which
\citet{herzog.etal2011} found to provide the best fit.
It included the log proportion of unvaccinated school starters
as a covariate in the autoregressive component.
This proportion can be regarded as a proxy for the fraction of
susceptibles, assuming that one dose of the MMR vaccine already provides
full protection.
The endemic component contained a sinusoidal wave~\eqref{eq:sincos} to
capture yearly seasonality,
and the standardized state population $n_{rt}$ as an offset. 
A neighbourhood effect was not included due to the coarse spatial resolution.
Indeed, suspected cases and their unprotected contacts are isolated
immediately \citep{who}, and the past week's incidence in adjacent states
is much less informative than within-state dynamics.
This model, called ``P0'' in the following, is given by 
\begin{align*}
Y_{rt}|\mathcal{F}_{t-1} &\sim \operatorname{Po}(\lambda_{rt} y_{r, t-1}+ \nu_{rt}),\\
\log(\lambda_{rt}) &= \alpha^{(\lambda)} + \beta^{(\lambda)} \log(1-x_{rt}),\\
\log(\nu_{rt}) &= \alpha^{(\nu)} + \delta \sin ( 2 \pi t/26 )
                 + \zeta \cos ( 2 \pi t/26 ) + \log(n_{rt}),
\end{align*}
where $x_{rt}$ is the estimated vaccination coverage for at least one dose.
This means that the local reproduction rate $\lambda_{rt}$ is
proportional to (a power of) the fraction of susceptibles,
which makes the epidemic component conform to the mass action principle.

\citet{Fisher2020} proposed an ecological Poisson model,
which also incorporates MMR vaccine effectiveness.
They used the conditional mean
\begin{equation*}
n_r (1 - \kappa x_r) (\lambda_r y_{r,t-1}/n_r + \nu_{rt}) ,
\end{equation*}
where $\kappa$ is the vaccine effect, $\lambda_r$ is the risk of infection 
and $\nu_{rt}$ is the endemic risk. 
To take the effect of the varying vaccination coverage into account, we
use a similar mean component in the following model extensions.
Moreover, we include both yearly and biennial seasonality in both endemic and epidemic components.
To account for overdispersion, we also switch to
negative binomial distributions
with state-specific overdispersion parameters.
This model, denoted by ``NB1'', is given by 
\begin{align*}
Y_{rt}|\mathcal{F}_{t-1} &\sim \operatorname{NB}(\lambda_{rt} y_{r, t-1}+ \nu_{rt}, \psi_r),\\
\log(\lambda_{rt}) &= \alpha^{(\lambda)}+ \sum_{s=1}^2 \left\{
    \delta_s^{(\lambda)} \sin \left( \tfrac{2 \pi t }{s \cdot 26} \right) + 
    \zeta_s^{(\lambda)} \cos \left( \tfrac{2 \pi t }{s \cdot 26} \right)
  \right\} + \log(1 - \kappa x_{rt}),\\
\log(\nu_{rt}) &= \alpha^{(\nu)} + \sum_{s=1}^2 \left\{
    \delta_s^{(\nu)} \sin \left( \tfrac{2 \pi t }{s \cdot 26} \right) + 
    \zeta_s^{(\nu)} \cos \left( \tfrac{2 \pi t }{s \cdot 26} \right)
  \right\} + \log((1 - \kappa x_{rt})\,n_{rt}) ,
\end{align*}
where $\kappa = 0.92$ is the posterior median estimated by \citet{Fisher2020},
and $s=1,2$ corresponds to yearly and biennial seasonality, respectively.
To account for unobserved heterogeneity between states, this model is
further extended with uncorrelated (``NB2'') or correlated (``NB3'') random
effects in both components.

Based on the structure of the models NB1 to NB3, we build several HHH4ZI
models of increasing complexity.
The simplest extension is ``ZI1'', which mixes NB1 with an
autoregressive zero-inflation probability $\gamma_{rt}$ via
\begin{equation*}
\logit(\gamma_{rt}) = \alpha^{(\gamma)} + \beta^{(\gamma)} y_{r,t-1} .
\end{equation*}
Models ``ZI2'' and ``ZI3'' are ZI1 models with additional uncorrelated
(ZI2) or correlated (ZI3) random effects in endemic, autoregressive and
zero-inflation components.
By incorporating yearly and biennial seasonality also in the
zero-inflation components of models ZI2 and ZI3, we obtain models ``ZI4''
and ``ZI5'', respectively.

\subsection{Results}

Table~\ref{tab:fit} summarizes the estimated parameters from the various models.
The estimated seasonal amplitudes in the autoregressive part are barely
affected by the model updates. The yearly pattern is stronger than the
biennial cycle; the combined effect with a peak in calendar weeks 11--12 and a
minimum at calendar weeks 39--40 is shown in Figure~S2 in the
supplementary material.
The estimated seasonal effect of the endemic component is very similar,
but shrinks when allowing for seasonality of the zero-inflation
probability (ZI4 and ZI5).
The large overdispersion $\hat\psi_r$ estimated for some states (Thuringia
and Bremen), is considerably reduced by accounting for state-specific
zero-inflation probabilities beyond the autoregressive effect (compare ZI1
to ZI2 or ZI3).
At the same time, the range of the time-varying effective reproduction number $R_t$
increases even more than after introducing state-specific overdispersion. This is
due to an improved fit of large outbreaks via the epidemic component
in states with otherwise low counts. 
We observe that $R_t$ is estimated to exceed the threshold 1 in January
and typically remains below one between June and December (see Figure~S3
in the supplementary material).
In 2015, $R_t$ even peaked at 3.1, which reflects a large measles outbreak
among refugees in Berlin \citep{Werber2017}.
We note that allowing for correlation between random effects
increases their variance (NB2 to NB3, ZI2 to ZI3, ZI4 to ZI5).
We discuss their spatial variation in the context of model ZI3 further below.

\begin{table}[ht]
\input{./tables/fit_sum1}
\input{./tables/fit_sum2}
\input{./tables/fit_sum3}
\caption{
Parameter estimates (and standard errors) and other summaries of model fit.
For region-specific parameters the table shows the range of the point estimates.
The time-varying dominant eigenvalue (maxEV) is also summarized by its range.
Estimates of $A_1^{\cdot}$ and $A_2^{\cdot}$ correspond to the amplitudes
of the yearly and biennial harmonics, respectively.
The $\hat\rho_{.,.}$ columns contain the estimated correlations between the
different random effect components.
For models with random effects, both the penalized log-likelihood $l_{\text{pen}}$ and
the marginal log-likelihood ($l_{\text{marg}}$) are given (but note that
these should not be used for model selection).
For models without random effects, $l_{\text{pen}}$ refers to the standard log-likelihood.
}
\label{tab:fit}
\end{table}

We assessed the quality of one-step-ahead forecasts during the last four years.
Several scores are considered \citep{czado.etal2009}:
logarithmic score (LS), 
Dawid-Sebastiani score (DSS), 
rank probability score (RPS), and 
squared error score (SES).
The logarithmic score is a strictly proper score, 
whereas the Dawid-Sebastiani score and the rank probability score  are proper. 
The squared error score is a classical measure of forecast performance. 
It is proper when it is taken as a score for probabilistic forecast \citep{gneiting.etal2007}. 
However it gives the same score to predictive distributions with the same expectation, 
regardless of the shape of distributions \citep{Broecker2007}. 
The maximum logarithmic score (maxLS) \citep{ray.etal2017} is not a proper
score, but is additionally used to evaluate the worst-case forecast performance.
Forecast performance using these scores is compared in Table~\ref{tab:forecast}.

\input{./tables/forecast_sum}

Models ZI3 and ZI5 consistently produce the best and second-best forecasts
in terms of the average LS, DSS and RPS.
Based on Monte Carlo permutation tests for differences in mean log scores,
model ZI3 significantly outperforms all other models except ZI5.
Adding the zero-inflation component consistently improves the
aforementioned scores (NB1 to ZI1, NB2 to ZI2 and NB3 to ZI3),
where the simple Poisson model ranks last.
However, the squared error score ranks models substantially different.
Model ZI1 ranks first followed by models NB3, NB1 and NB2.
Model ZI5 has the worst mean squared error score.
The root mean squared prediction errors of all models are around 4,
which means that the forecasts differ from the observed counts
by around four cases on average.
Looking at the maximum logarithmic score, 
models have again different ranks.
Model NB1 has the best (i.e., lowest) maximum log score,
followed by models ZI1, ZI4, ZI5, ZI3 and ZI2.
The Poisson model performs much worse at times and on average.

\begin{figure}[h]
\centering
\includegraphics[width=0.32\textwidth]{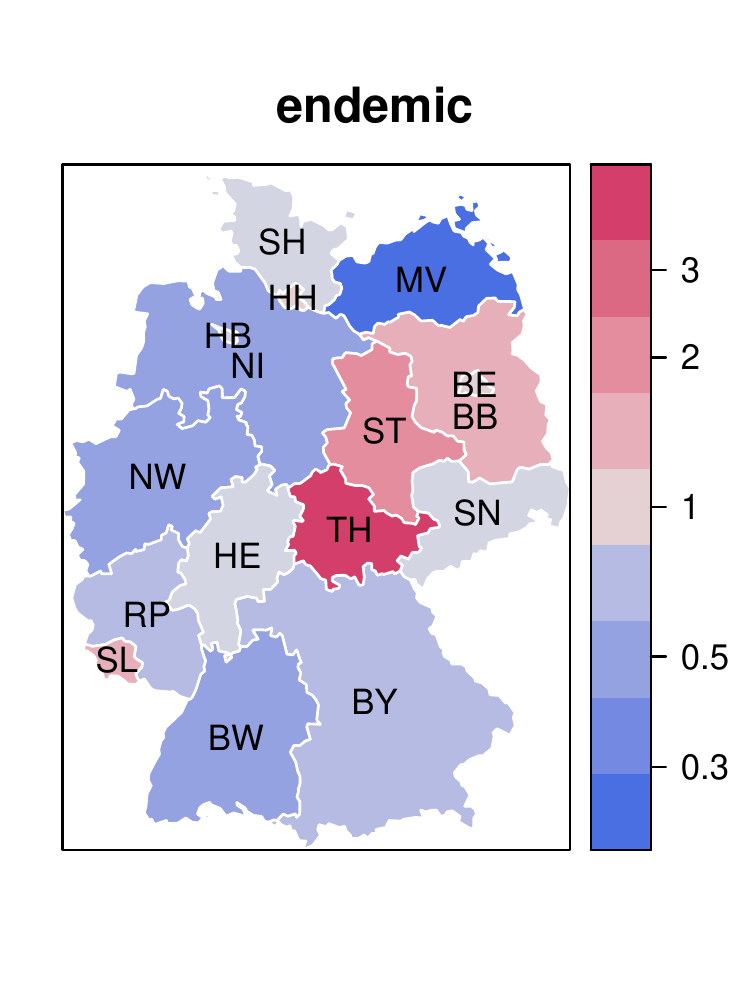}
\includegraphics[width=0.32\textwidth]{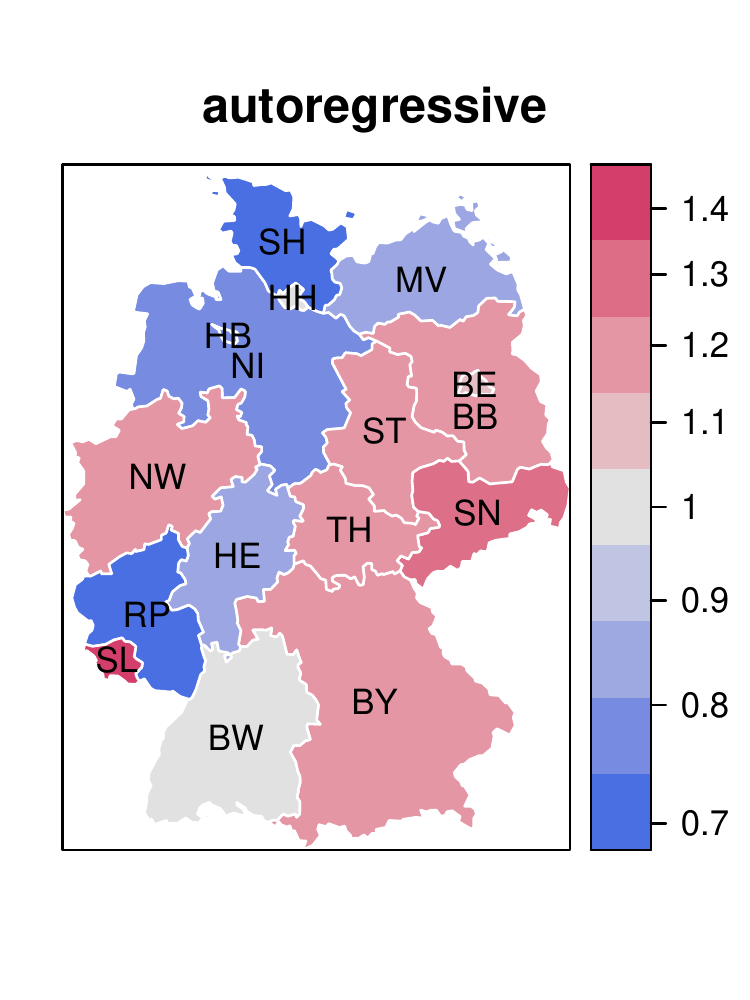}
\includegraphics[width=0.32\textwidth]{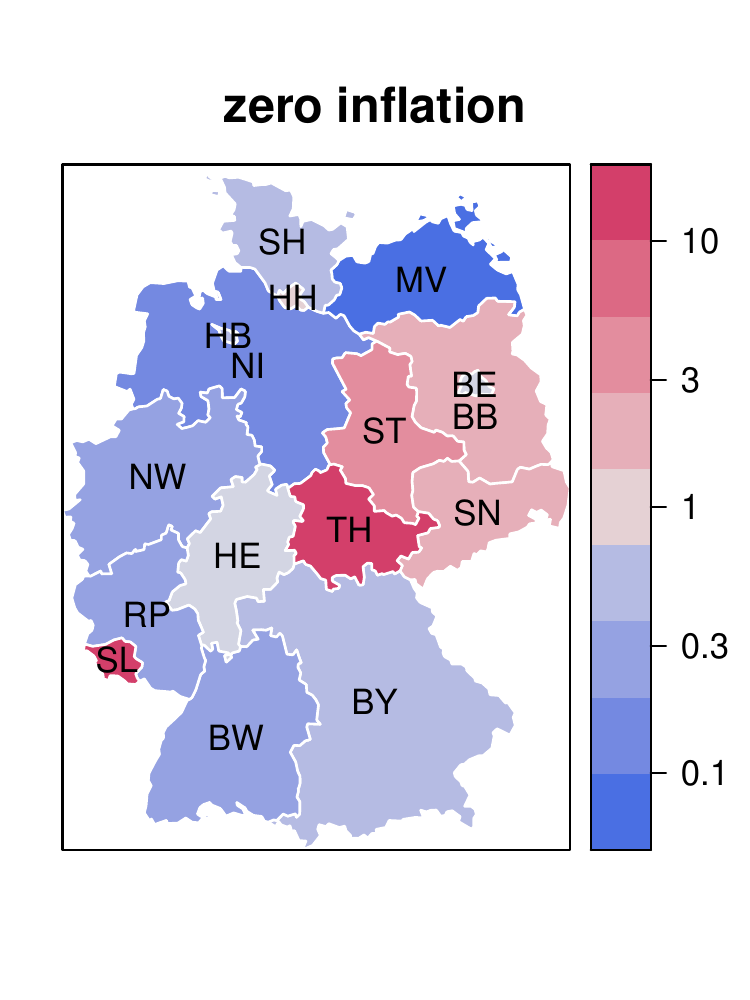}
\caption{Maps of exp-transformed region-specific intercepts estimated in
  model ZI3. These correspond to rate ratios (left, center) and odds
  ratios (right), respectively.}
  \label{fig:ZI3_ri}
\end{figure}

We now focus on model ZI3 and discuss the remaining model parameters.
The estimated autoregressive parameter 
in the zero-inflation component is $\hat\beta^{(\gamma)} = -0.82$
(95\% CI: [-1.07, -0.57]), which means that each additional case at time
$t-1$ decreases the odds of an excess zero at time $t$
by $1 - \exp(-0.82) = 56\%$.
Plots of the fitted time series in Figure~S4 (supplementary
material) show that model ZI3 fits reasonably well, especially in
Brandenburg, Hamburg, Saarland, Saxony-Anhalt and Thuringia.
Figure~\ref{fig:ZI3_ri} shows maps of the exp-transformed
region-specific intercepts estimated from model ZI3.
For the endemic and autoregressive components these
correspond to rate ratios taking the regional average as a reference, 
whereas for the zero-inflation component the values correspond to odds ratios.
Five states have a relatively high zero-inflation (odds ratio larger than 3):
Brandenburg (BB), Saarland (SL), Saxony (SN), Saxony-Anhalt (ST) and Thuringia (TH).
Except for Saarland (SL), these states are in East Germany.
We notice that these five states have a relatively high vaccination coverage, 
a low total number of cases (Table~\ref{tab:fact}), but
a few outbreaks (Figure~\ref{fig:measles_ts}).
This pattern can be accomodated by increasing both the zero-inflation
probability and the autoregressive parameter. Correspondingly, these five
states also have the largest autoregressive intercepts (rate ratios above 1.18). 
In contrast, Lower Saxony (NI) and Mecklenburg-Western Pomerania (MV) have 
the lowest zero inflation. This may relate to a lack of large outbreaks
with consistently low counts that would even conform to a negative
binomial distribution.

\section{Conclusion}

We have proposed a multivariate zero-inflated endemic-epidemic model for infectious disease counts with excessive zeros. 
This model consists of two parts: a zero-inflation part, 
which represents the not-at-risk population,
and a HHH part, which models the at-risk population.
The zero-inflation part can incorporate seasonality and autoregressive terms. 
Its exp-transformed parameters can be interpreted as odds ratios.
The HHH part has the same endemic-epidemic decomposition as the
classical HHH framework.
Random effects are allowed in both parts to account for heterogeneity across units. 
We applied this model to state-level measles counts in Germany.
Both yearly and biennial seasonality are included in our models
to capture a potential biennial cycle of measles epidemics.
The effect of spatio-temporally varying vaccination coverage is accounted
for by assuming the incidence to be proportional to the unvaccinated
proportion of school starters.
We assessed the forecast performance of this model using proper scoring rules.
In comparison with negative binomial models with the same HHH part, 
zero-inflated HHH models can better capture time series where both
low-incidence periods and large outbreaks occur,
and the extended models also consistently improve forecast performance.


\subsection*{Conflict of interest}

The authors have declared no conflict of interest.

\subsection*{Funding}

This work was financially supported by the Interdisciplinary Center for
Clinical Research (IZKF) of the Friedrich-Alexander-Universit\"at
Erlangen-N\"urnberg (FAU), Germany [project J75].
Junyi Lu performed the present work in partial fulfilment of the
requirements for obtaining the degree 'Dr. rer. biol. hum.' at the FAU.

\subsection*{Data availability statement}

The data that support the findings of this study were derived from the
following resources available in the public domain:
case counts from SurvStat@RKI 2.0, Robert Koch Institute
(\url{https://survstat.rki.de}, accessed 26 March 2021),
vaccination coverage from the Information System of the Federal Health
Monitoring (\url{https://www.gbe-bund.de/}, accessed 23 February 2021),
and population data from the Federal Statistical Office of Germany
(Statistisches Bundesamt, \url{https://www.destatis.de/}, accessed 10 April 2021).
The derived datasets are part of the dedicated R package at
\url{https://github.com/Junyi-L/hhh4ZI}.
Code to reproduce all results using that package is provided in the
supplementary material.

\subsection*{ORCID}

\textit{Sebastian Meyer} \url{https://orcid.org/0000-0002-1791-9449}

\renewcommand{\bibfont}{\small}
\bibliographystyle{apalike}
\bibliography{references}

\section*{Appendix}

\subsection*{Parametrization of $\bm{\Sigma}$}

When assuming within-unit correlation of the different random effects,
we use a spherical parametrization \citep{Pinheiro1996, Rapisarda2007} for
the matrix $\bm{\Omega}$.
This enforces positive definiteness and ensures computational efficiency.
 
We first factorize the covariance matrix $\bm{\Omega}$ by Cholesky decomposition
$
\bm{\Omega} = \mathbf{D} \mathbf{L} \mathbf{L}^T \mathbf{D},
$
where $\mathbf{D} = \text{diag}(\sigma_1, \sigma_2, \sigma_3, \sigma_4)$ 
is a diagonal matrix with standard deviations (which we estimate on the log scale)
and $\mathbf{L} \in \mathbb{R}^{4 \times 4}$ is a lower triangular matrix.
To ensure the positive definiteness of the covariance matrix, 
the matrix $\mathbf{L}$ is parametrized by \citep{Rapisarda2007}
\begin{equation*}
\mathbf{L} = 
\begin{pmatrix}
1  \\
\frac{r_1}{\sqrt{r_1^2 + 1}} & \frac{1}{\sqrt{r_1^2 + 1}}  \\
\frac{r_2}{\sqrt{r_2^2 + 1}} & 
\frac{r_3}{\sqrt{r_3^2 + 1}\sqrt{r_2^2 + 1}} &
 \frac{1}{\sqrt{r_3^2 + 1}\sqrt{r_2^2 + 1}}  \\
\frac{r_4}{\sqrt{r_4^2 + 1}} & 
\frac{r_5}{\sqrt{r_5^2 + 1}\sqrt{r_4^2 + 1}} & 
\frac{r_6}{\sqrt{r_6^2 + 1}\sqrt{r_5^2 + 1}\sqrt{r_4^2 + 1}}& 
\frac{1}{\sqrt{r_6^2 + 1}\sqrt{r_5^2 + 1}\sqrt{r_4^2 + 1}}
\end{pmatrix},
\end{equation*}
where $r_1$, $r_2$, $r_3$, $r_4$, $r_5$ and $r_6$ are unconstrained parameters.

The log-determinant of $\bm{\Sigma}$, which appears in the marginal log-likelihood $l_{\text{marg}}$,
is given by
\begin{equation*}
\log |\bm{\Sigma}| = 2R\left(\sum_{i = 1}^4 \log\sigma_i - \frac{1}{2} \sum_{i = 1}^6 \log(r_i^2 + 1)\right).
\end{equation*}

\subsection*{Score function}
The score function and Fisher information matrix are derived 
based on the score function and Fisher information matrix of 
the HHH model, which can be found in \citet{paul.held2011}.
We denote the score function of the penalized log-likelihood with respect to the fixed and random parameters by 
$\mathbf{s}_{\text{pen}}(\bm{\theta},\bm{b},\bm{\tilde{\psi}}; \bm{\Sigma})$. 
It can be partitioned as 
\begin{equation*}
\bm{s}_{\text{pen}}(\bm{\theta},\bm{b},\bm{\tilde{\psi}}; \bm{\Sigma}) 
=
\begin{pmatrix}
\bm{s}(\bm{\theta}) \\
\bm{s}(\bm{\tilde{\psi}})\\
\bm{s}(\bm{b}) - \bm{\Sigma}^{-1}\bm{b}
\end{pmatrix},
\end{equation*}
where $\bm{s}(\bm{i})=\frac{\partial l(\bm{\theta}, \bm{b}, \bm{\tilde{\psi}})}{\partial \bm{i}}$ 
corresponds to the unpenalized score vector with respect to the parameter vector $\bm{i}$.
We define $\bm{\xi}^{(\nu)} = (\alpha^{(\nu)}, \bm{\beta}^{(\nu)T}, \bm{b}^{(\nu)T)})^T$ 
as the vector of parameters of fixed and random effects in the endemic ($\nu$) component. 
Vectors $\bm{\xi}^{(\lambda)}$, $\bm{\xi}^{(\phi)}$ and $\bm{\xi}^{(\gamma)}$ are defined analogously.
We denote by $l_{rt}(\bm{\theta}, \bm{b}, \tilde\psi_r) =
\log f_{ZI}(y_{rt}; \mu_{rt}, \psi_r, \gamma_{rt})$ the terms of the
unpenalized log-likelihood of the zero-inflated model, and by
$l_{rt}^H(\mu_{rt}, \tilde{\psi}_r) = \log f(y_{rt}; \mu_{rt}, \psi_r)$ 
the terms of the unpenalized log-likelihood of the original HHH model.

For $\bm{i} = \bm{\xi}^{(\nu)}$, $\bm{\xi}^{(\lambda)}$, $\bm{\xi}^{(\phi)}$, $\bm{\tilde{\psi}}$,
\begin{equation*}
\bm{s}(\bm{i}) =
\sum_{rt} \exp(-l_{rt}(\bm{\theta}, \bm{b}, \tilde\psi_r)) (1 - \gamma_{rt})
f(y_{rt}, \mu_{rt}, \psi_r) 
\frac{\partial l^H_{rt}(\mu_{rt}, \tilde{\psi}_r)}{\partial \bm{i}},
\end{equation*}
and
\begin{equation*}
\bm{s}(\bm{\xi}^{(\gamma)})  =
\sum_{rt} \exp(-l_{rt}(\bm{\theta}, \bm{b}, \tilde\psi_r))(\indy - f(y_{rt}, \mu_{rt}, \psi_r) )\frac{\partial \gamma_{rt}}{\bm{\xi}^{\gamma}}.
\end{equation*}

We denote $g(x) = \frac{\exp(x)}{\exp(x) + 1}$, and 
$g'(x)  = \frac{\exp(x)}{(\exp(x) + 1)^2}$,
$g''(x) = \frac{\exp(x)(1 - \exp(x))}{(\exp(x) + 1)^3}$.
Then $\gamma_{rt} = g(\bm{\xi}^{(\gamma)T} \bm{u}_{rt}^{(\gamma)})$,
where $\bm{u}_{rt}^{(\gamma)} = (1, \bm{x}_{rt}^{(\gamma)T}, \bm{z}_{rt}^{(\gamma)T})^T$, and $\bm{z}_{rt}^{(\gamma)T}$ is a unit vector with $r$-th element being 1.
We then have 
$\frac{\partial \gamma_{rt}}{\partial \bm{\xi}^{(\gamma)}} = g'(\bm{\xi}^{(\gamma)T} \bm{u}_{rt}^{(\gamma)})\bm{u}_{rt}^{(\gamma)}$,
and
$\frac{\partial^2 \gamma_{rt}}{\partial \bm{\xi}^{(\gamma)} \partial \bm{\xi}^{(\gamma)T}} = g''(\bm{\xi}^{(\gamma)T} \bm{u}_{rt}^{(\gamma)})\bm{u}_{rt}^{(\gamma)}\bm{u}_{rt}^{(\gamma)^T}.
$

\subsection*{Fisher information matrix}

The observed Fisher information matrix can be partitioned as 
\begin{equation*}
\Fpen(\bm{\theta},\bm{b},\bm{\tilde{\psi}}; \bm{\Sigma})
= 
\begin{pmatrix}
\Fpen[\bm{\theta},\bm{\theta}]
& \Fpen[\bm{\theta},\bm{b}]
& \Fpen[\bm{\theta},\bm{\tilde{\psi}}] \\
\Fpen[\bm{b},\bm{\theta}]
& \Fpen[\bm{b},\bm{b}] + \bm{\Sigma}^{-1}
& \Fpen[\bm{b},\bm{\tilde{\psi}}] \\
\Fpen[\bm{\tilde{\psi}},\bm{\theta}]
& \Fpen[\bm{\tilde{\psi}},\bm{b}]
& \Fpen[\bm{\tilde{\psi}},\bm{\tilde{\psi}}]
\end{pmatrix},
\end{equation*}
where $\Fpen[\bm{i},\bm{j}] =- \frac{\partial^2 l(\bm{\theta}, \bm{b}, \bm{\tilde{\psi}})}{\partial \bm{i} \partial \bm{j}^T}$ 
denotes the block of the unpenalized Fisher information matrix corresponding to the parameter vectors $\bm{i}$ and $\bm{j}$.
For $\bm{i}, \bm{j} = \bm{\xi}^{(\lambda)}$, $\bm{\xi}^{(\phi)}$, $\bm{\xi}^{(\nu)}$, or $\bm{\tilde{\psi}}$,
\begin{align*}
\Fpen[\bm{i}, \bm{j}] &= 
\Fpen[\bm{j}, \bm{i}] = -\sum_{rt}
(1 - \gamma_{rt})
f(y_{rt}, \mu_{rt}, \psi_r) 
\exp(-l_{rt}(\bm{\theta}, \bm{b}, \tilde\psi_r))
\\
& \bigg[-\exp(-l_{rt}(\bm{\theta}, \bm{b}, \tilde\psi_r)) (1 - \gamma_{rt})
f(y_{rt}, \mu_{rt}, \psi_r)
 \frac{\partial l^H_{rt}(\mu_{rt},\tilde{\psi}_r)}{\partial \bm{j}}
(\frac{\partial l^H_{rt}(\mu_{rt}, \tilde{\psi}_r)}{\partial \bm{i}})^T \\
&  \quad +
\frac{\partial l^H_{rt}(\mu_{rt}, \tilde{\psi}_r)}{\partial \bm{j}}
(\frac{\partial l^H_{rt}(\mu_{rt}, \tilde{\psi}_r)}{\partial \bm{i}})^T  
+
\frac{\partial^2 l^H_{rt}(\mu_{rt}, \tilde{\psi}_r)}{\partial \bm{i} \partial \bm{j}^T}
\bigg],
\end{align*}
\begin{align*}
\Fpen[\bm{i}, \bm{\xi}^{(\gamma)}] &= \Fpen[\bm{\xi}^{(\gamma)}, \bm{i}] = -\sum_{rt}
f(y_{rt}, \mu_{rt}, \psi_r)
\frac{\partial l^H_{rt}(\mu_{rt}, \tilde{\psi}_r)}{\partial \bm{i}} \exp(-l_{rt}(\bm{\theta}, \bm{b}, \tilde\psi_r))\\
& \bigg[-\exp(-l_{rt}(\bm{\theta}, \bm{b}, \tilde\psi_r))
 (\indy -  f(y_{rt}, \mu_{rt}, \psi_r))
\frac{\partial \gamma_{rt}}{\partial \bm{\xi}^{(\gamma)}}
(1 - \gamma_{rt})
- \frac{\partial \gamma_{rt}}{\partial \bm{\xi}^{(\gamma)}}
\bigg],
\end{align*}
\begin{align*}
\Fpen[\bm{\xi}^{(\gamma)}, \bm{\xi}^{(\gamma)}]
&=-\sum_{rt}
 (\indy -  f(y_{rt}, \mu_{rt}, \psi_r)) \exp(-l_{rt}(\bm{\theta}, \bm{b}, \tilde\psi_r))\\
& \bigg[-\exp(-l_{rt}(\bm{\theta}, \bm{b}, \tilde\psi_r))
 (\indy -  f(y_{rt}, \mu_{rt}, \psi_r))
\frac{\partial \gamma_{rt}}{\partial \bm{\xi}^{(\gamma)}}
\frac{\partial \gamma_{rt}}{\partial \bm{\xi}^{(\gamma)T}}
 + \frac{\partial^2 \gamma_{rt}}{\partial \bm{\xi}^{(\gamma)} \bm{\xi}^{(\gamma)T}}
\bigg]
.
\end{align*}

\end{document}



\maketitle

\listoffigures

\newpage

\begin{figure}
\centering
\includegraphics[width=\textwidth]{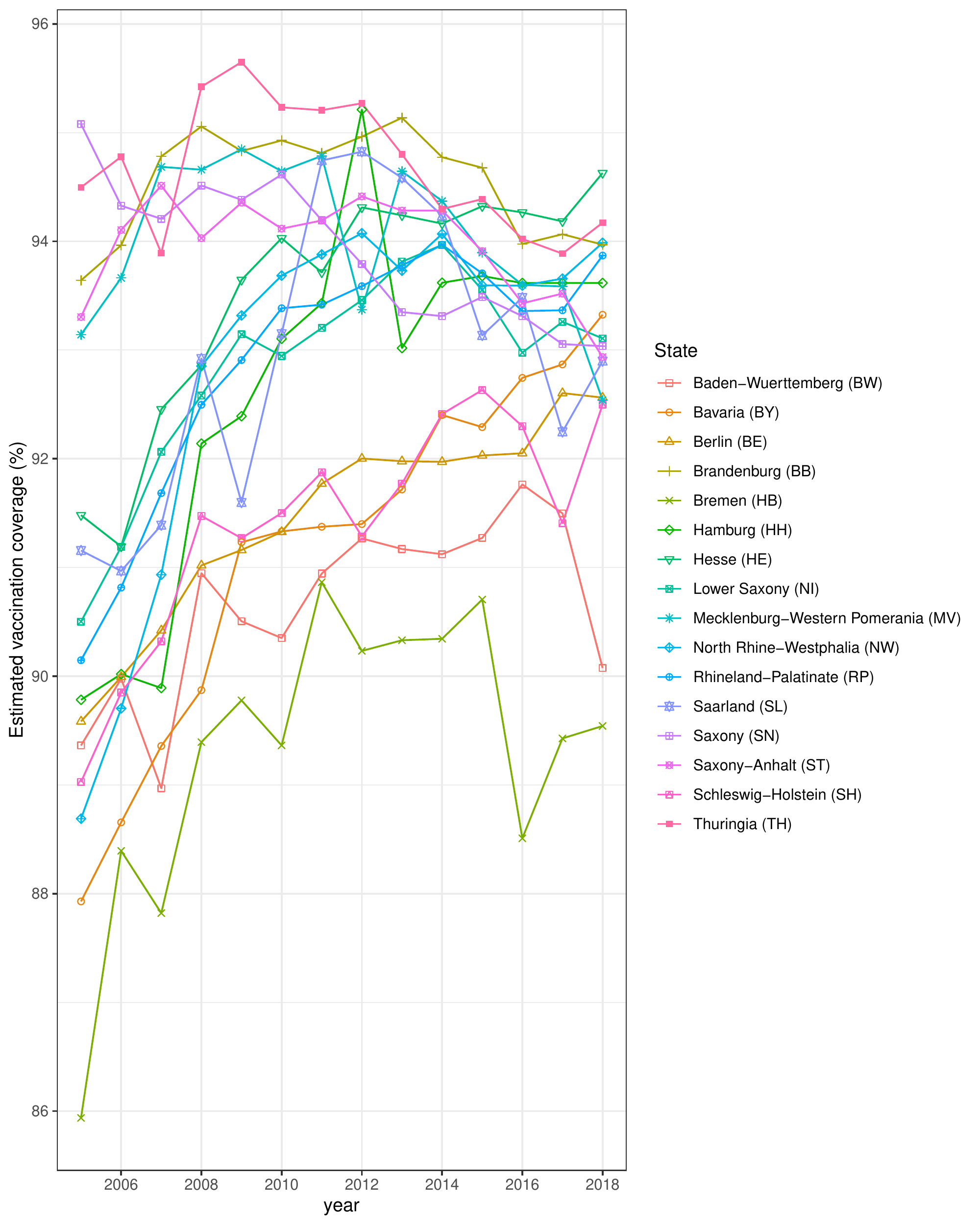}
\caption{Measles-mumps-rubella (MMR) vaccination coverage rates of
  children at school entry, by federal state in Germany, 2005--2018.}
\end{figure}


\begin{figure}
\centering
\includegraphics[width=\textwidth]{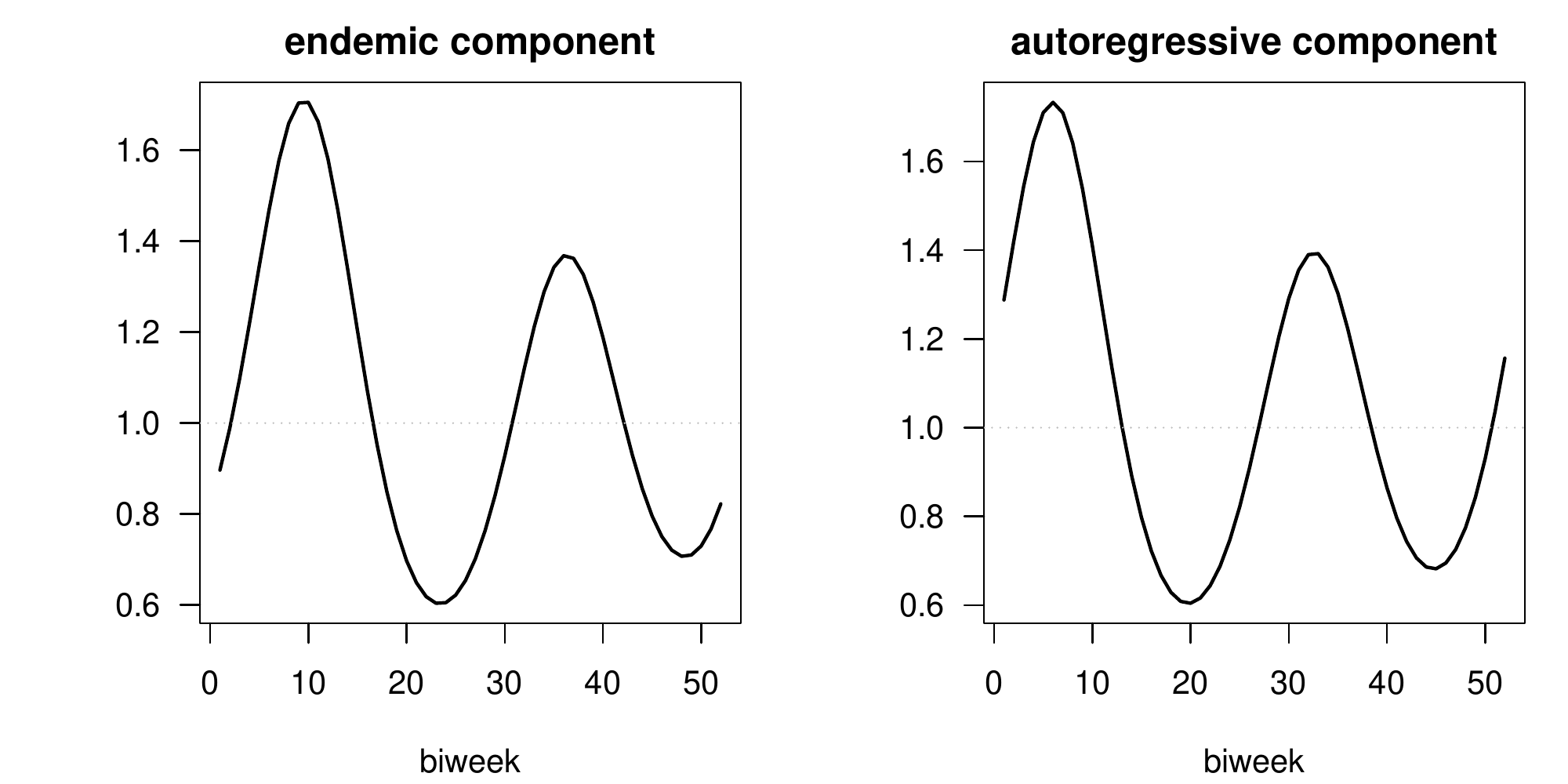}
\caption{Estimated biennial seasonality of the endemic and autoregressive components of model ZI3.}
\end{figure}

\begin{figure}
\centering
\includegraphics[width=\textwidth]{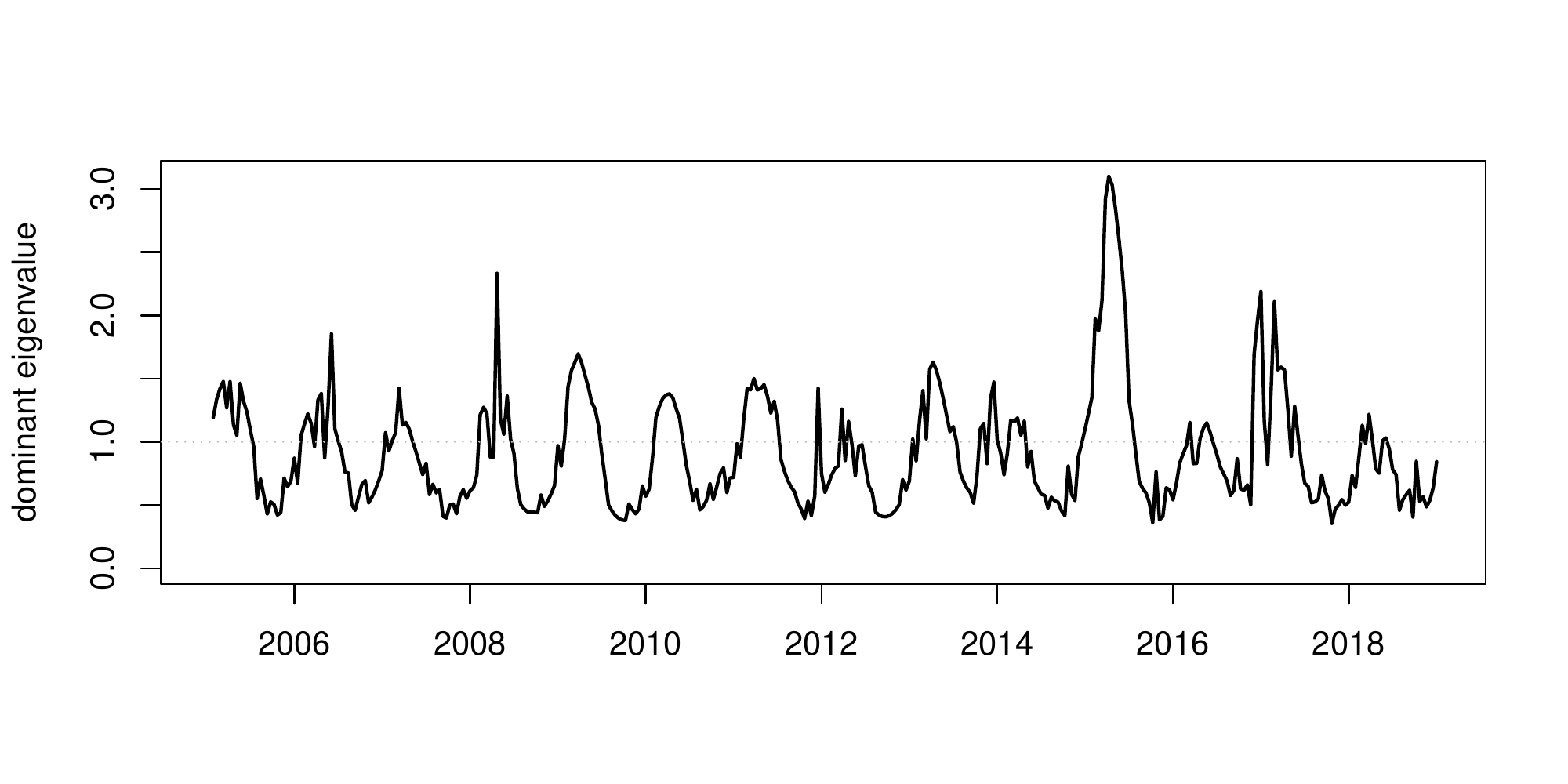}
\caption{Time-varying effective reproduction number $R_t$ as estimated by model ZI3.}
\end{figure}

\begin{figure}
\centering
\includegraphics[width=\textwidth]{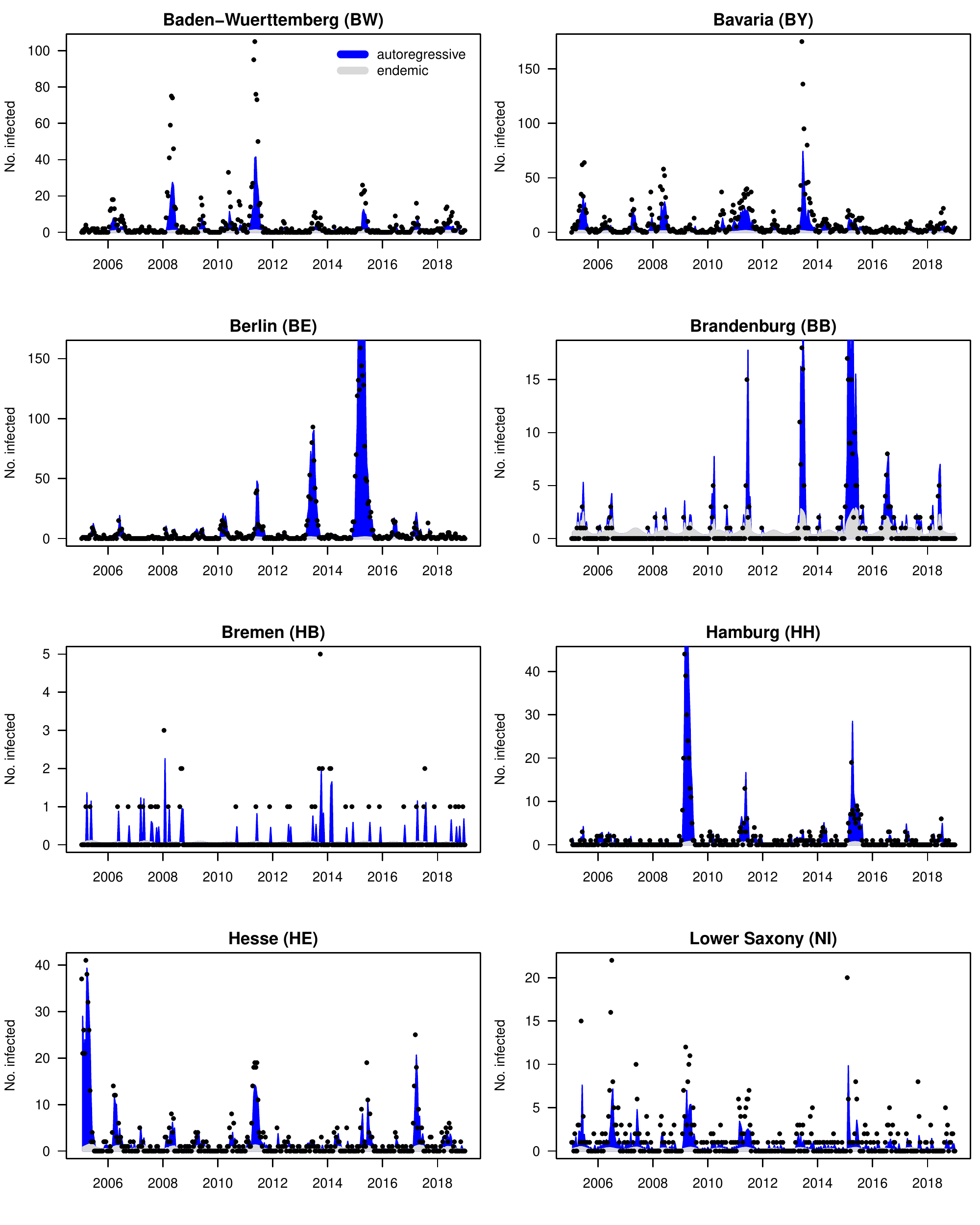}
\emph{(continued on next page)}
\end{figure}

\begin{figure}
\centering
\includegraphics[width=\textwidth]{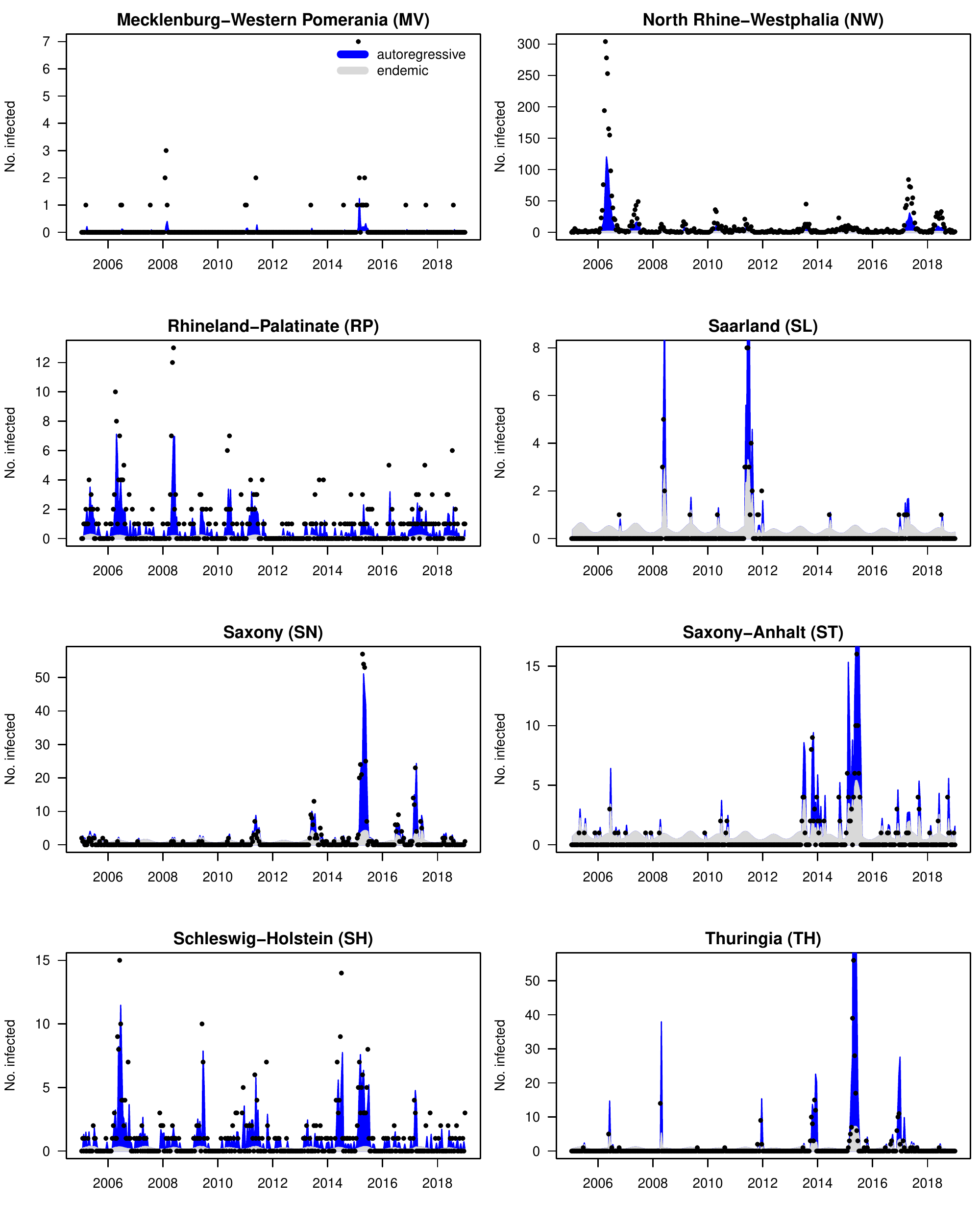}
\caption{Fitted mean of model ZI3.}
\end{figure}

\begin{figure}
\centering
\includegraphics[width=\textwidth]{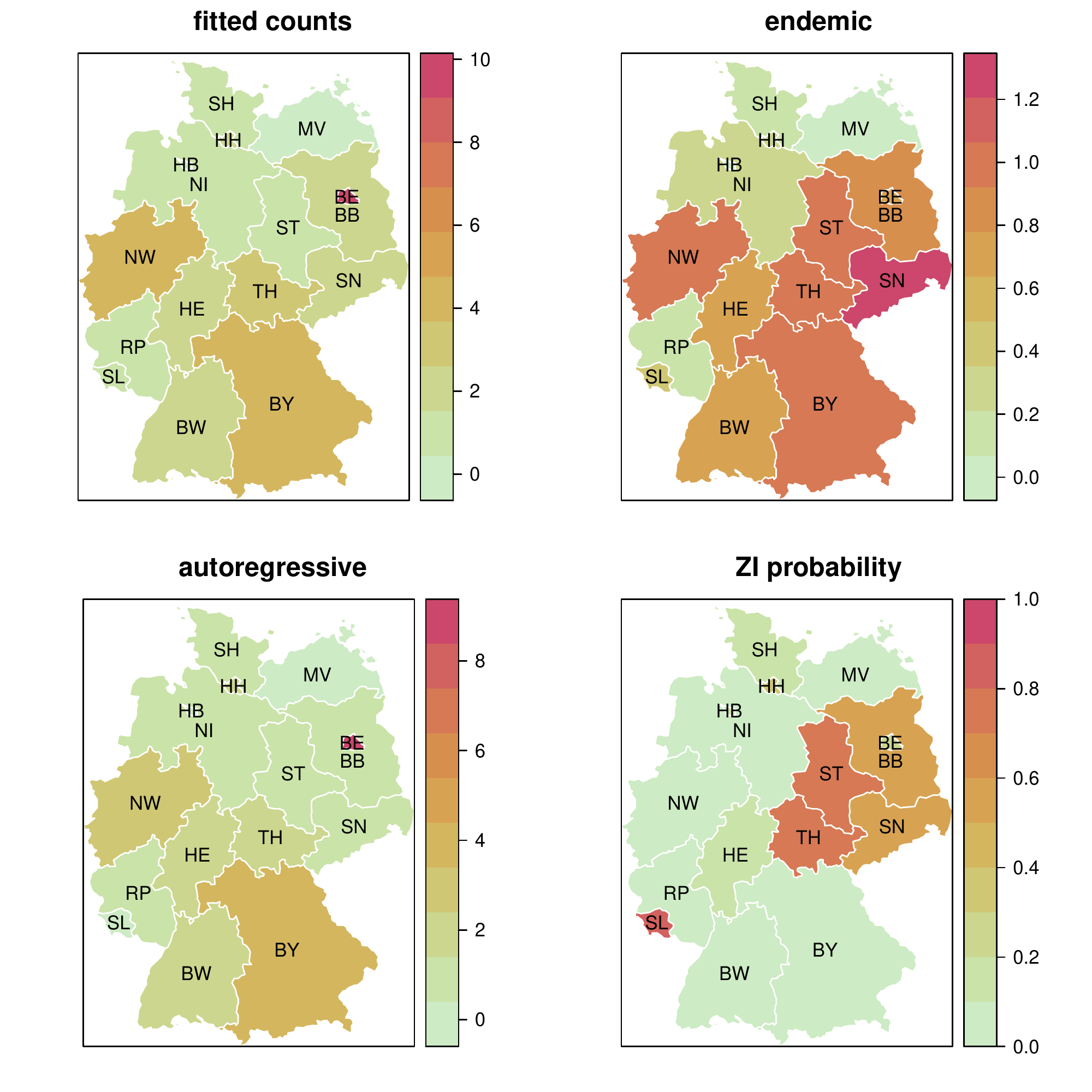}
\caption{Maps of fitted components in model ZI3, averaged over the whole time period.}
\end{figure}


%% file: tables/sim.tex
\begin{table}[ht]
\centering
\begin{tabular}{ll|rrrrrrrr}
  \hline
$T$ &  & $\hat{\alpha}^{(\lambda)}$ & $\hat{\alpha}^{(\phi)}$ & $\hat{\alpha}^{(\nu)}$ & $\hat{\alpha}^{(\gamma)}$ & $\hat{\beta}^{(\gamma)}_1$ & $\hat{\beta}^{(\gamma)}_2$ & $\hat{\beta}^{(\gamma)}_3$ & $\hat{\psi}$ \\ 
  \hline
50 & mean & -0.32 & -0.13 & 0.50 & 0.20 & 0.40 & -0.31 & -0.10 & 0.49 \\ 
   & SD & 0.12 & 2.79 & 0.10 & 0.13 & 0.13 & 0.14 & 0.03 & 0.10 \\ 
   & coverage 95 & 95.00 & 95.60 & 94.70 & 95.50 & 95.10 & 94.70 & 94.80 & 93.40 \\ 
   & coverage 50 & 49.90 & 55.40 & 51.80 & 52.10 & 50.80 & 49.10 & 49.10 & 49.90 \\ 
   \hline
100 & mean & -0.32 & 0.43 & 0.50 & 0.20 & 0.40 & -0.30 & -0.10 & 0.50 \\ 
   & SD & 0.08 & 0.39 & 0.07 & 0.09 & 0.09 & 0.09 & 0.02 & 0.07 \\ 
   & coverage 95 & 95.10 & 96.60 & 95.90 & 95.60 & 94.90 & 94.90 & 94.80 & 94.70 \\ 
   & coverage 50 & 49.50 & 52.90 & 51.40 & 47.30 & 51.80 & 52.50 & 50.50 & 50.70 \\ 
   \hline
500 & mean & -0.30 & 0.49 & 0.50 & 0.20 & 0.40 & -0.30 & -0.10 & 0.50 \\ 
   & SD & 0.04 & 0.14 & 0.03 & 0.04 & 0.04 & 0.04 & 0.01 & 0.03 \\ 
   & coverage 95 & 95.90 & 96.40 & 94.40 & 94.30 & 95.60 & 94.90 & 95.30 & 94.40 \\ 
   & coverage 50 & 50.50 & 50.80 & 51.80 & 48.10 & 49.80 & 49.30 & 49.20 & 48.00 \\ 
   \hline
\end{tabular}
\caption{Means and standard deviations (SD) of parameter estimates
             from 1000 simulations for various time series lengths ($T$).
             Coverage probabilities of 95\% and 50\% confidence intervals are also given.} 
\label{tab:sim}
\end{table}

%% file: tables/fact.tex
\begin{table}[ht]
\centering
\begin{tabular}{l|rrrrr}
  \hline
State & Max & Zeros & Total & Population & Coverage \\ 
  \hline
Baden-Wuerttemberg (BW) & 105 & 147 & 1\,733 & 10\,773\,318 & 89\%\,\dots 90\% \\ 
  Bavaria (BY) & 175 & 66 & 3\,082 & 12\,654\,101 & 88\%\,\dots 93\% \\ 
  Berlin (BE) & 159 & 151 & 2\,533 & 3\,464\,046 & 90\%\,\dots 93\% \\ 
  Brandenburg (BB) & 18 & 276 & 309 & 2\,498\,962 & 94\%\,\dots 94\% \\ 
  Bremen (HB) & 5 & 324 & 53 & 665\,379 & 86\%\,\dots 90\% \\ 
  Hamburg (HH) & 44 & 233 & 498 & 1\,773\,744 & 90\%\,\dots 94\% \\ 
  Hesse (HE) & 41 & 181 & 859 & 6\,105\,864 & 91\%\,\dots 95\% \\ 
  Lower Saxony (NI) & 22 & 181 & 504 & 7\,909\,337 & 91\%\,\dots 93\% \\ 
  Mecklenburg-Western Pomerania (MV) & 7 & 340 & 36 & 1\,634\,664 & 93\%\,\dots 93\% \\ 
  North Rhine-Westphalia (NW) & 304 & 100 & 3\,641 & 17\,831\,720 & 89\%\,\dots 94\% \\ 
  Rhineland-Palatinate (RP) & 13 & 203 & 334 & 4\,033\,268 & 90\%\,\dots 94\% \\ 
  Saarland (SL) & 8 & 341 & 55 & 1\,010\,670 & 91\%\,\dots 93\% \\ 
  Saxony (SN) & 57 & 278 & 495 & 4\,127\,619 & 95\%\,\dots 93\% \\ 
  Saxony-Anhalt (ST) & 16 & 303 & 173 & 2\,309\,032 & 93\%\,\dots 93\% \\ 
  Schleswig-Holstein (SH) & 15 & 213 & 355 & 2\,841\,994 & 89\%\,\dots 92\% \\ 
  Thuringia (TH) & 56 & 321 & 302 & 2\,212\,890 & 94\%\,\dots 94\% \\ 
   \hline
\end{tabular}
\caption{Maximum count, number of zeros, total of
bi-weekly measles counts, (average) population, and
estimated vaccination coverage in 2005 and 2018,
for the 16 federal states of Germany.} 
\label{tab:fact}
\end{table}

%% file: tables/fit_sum1.tex
\begin{tabular}{lllllllll}
  \hline
Model & $\hat{\alpha}^{(\lambda)}$ & $\hat{\sigma}_{\lambda}$ & $\hat{A}_1^{(\lambda)}$ & $\hat{A}_2^{(\lambda)}$ & $\hat{\alpha}^{(\nu)}$ & $\hat{\sigma}_{\nu}$ & $\hat{A}_1^{(\nu)}$ & $\hat{A}_2^{(\nu)}$ \\ 
  \hline
P0 & 0.71 (0.12) &  &  &  & 1.92 (0.03) &  & 0.55 (0.04) &  \\ 
  NB1 & 1.38 (0.04) &  & 0.43 (0.05) & 0.11 (0.06) & 3.98 (0.03) &  & 0.44 (0.05) & 0.13 (0.05) \\ 
  NB2 & 1.01 -- 2.62 & 0.22 & 0.44 (0.05) & 0.13 (0.06) & 3.01 -- 7.88 & 0.48 & 0.44 (0.05) & 0.13 (0.05) \\ 
  NB3 & 1.00 -- 2.65 & 0.23 & 0.44 (0.05) & 0.13 (0.06) & 3.03 -- 7.88 & 0.48 & 0.44 (0.05) & 0.13 (0.05) \\ 
  ZI1 & 1.40 (0.04) &  & 0.42 (0.05) & 0.10 (0.06) & 4.23 (0.05) &  & 0.41 (0.05) & 0.12 (0.05) \\ 
  ZI2 & 0.99 -- 2.68 & 0.24 & 0.44 (0.05) & 0.13 (0.06) & 3.68 -- 9.11 & 0.63 & 0.43 (0.05) & 0.13 (0.05) \\ 
  ZI3 & 0.96 -- 2.69 & 0.27 & 0.44 (0.05) & 0.12 (0.06) & 2.96 -- 9.10 & 0.79 & 0.42 (0.05) & 0.14 (0.05) \\ 
  ZI4 & 0.99 -- 2.71 & 0.24 & 0.43 (0.05) & 0.12 (0.06) & 3.78 -- 9.14 & 0.58 & 0.37 (0.06) & 0.06 (0.06) \\ 
  ZI5 & 0.97 -- 2.78 & 0.28 & 0.43 (0.05) & 0.12 (0.06) & 3.76 -- 9.23 & 0.66 & 0.38 (0.06) & 0.08 (0.06) \\ 
   \hline
\end{tabular}

%% file: tables/fit_sum2.tex
\begin{tabular}{lllllll}
  \hline
Model & $\hat{\alpha}^{(\gamma)}$ & $\hat{\sigma}_{\gamma}$ & $\hat{\beta}^{(\gamma)}$ & $\hat{A}_1^{(\gamma)}$ & $\hat{A}_2^{(\gamma)}$ & $\hat{\psi}_r$ \\ 
  \hline
P0 &  &  &  &  &  &  \\ 
  NB1 &  &  &  &  &  & 0.49 -- 10.27 \\ 
  NB2 &  &  &  &  &  & 0.47 -- 10.30 \\ 
  NB3 &  &  &  &  &  & 0.47 -- 10.30 \\ 
  ZI1 & -1.17 (0.18) &  & -0.49 (0.09) &  &  & 0.41 -- 8.03 \\ 
  ZI2 & -1.95 -- 1.45 & 1.25 & -0.73 (0.12) &  &  & 0.45 -- 2.84 \\ 
  ZI3 & -3.79 -- 1.82 & 1.77 & -0.82 (0.13) &  &  & 0.31 -- 3.38 \\ 
  ZI4 & -2.01 -- 1.41 & 1.16 & -0.66 (0.11) & 0.32 (0.14) & 0.26 (0.13) & 0.44 -- 2.54 \\ 
  ZI5 & -1.74 -- 1.70 & 1.37 & -0.72 (0.11) & 0.26 (0.14) & 0.21 (0.14) & 0.39 -- 3.02 \\ 
   \hline
\end{tabular}

%% file: tables/fit_sum3.tex
\begin{tabular}{lllllrr}
  \hline
Model & $\hat{\rho}_{\lambda,\nu}$ & $\hat{\rho}_{\nu,\gamma}$ & $\hat{\rho}_{\lambda,\gamma}$ & maxEV & $l_{\text{pen}}$ & $l_{\text{marg}}$ \\ 
  \hline
P0 &  &  &  & 0.86 -- 1.00 & -9299.52 &  \\ 
  NB1 &  &  &  & 0.41 -- 1.42 & -7059.04 &  \\ 
  NB2 &  &  &  & 0.40 -- 1.52 & -6966.87 & -92.60 \\ 
  NB3 & -0.50 &  &  & 0.40 -- 1.52 & -6966.51 & -91.70 \\ 
  ZI1 &  &  &  & 0.34 -- 1.28 & -7037.44 &  \\ 
  ZI2 &  &  &  & 0.37 -- 2.42 & -6914.28 & -118.74 \\ 
  ZI3 & 0.30 & 0.86 & 0.62 & 0.36 -- 3.10 & -6909.21 & -111.99 \\ 
  ZI4 &  &  &  & 0.36 -- 2.23 & -6911.89 & -125.90 \\ 
  ZI5 & 0.22 & 0.70 & 0.65 & 0.34 -- 2.82 & -6908.55 & -121.53 \\ 
   \hline
\end{tabular}

%% file: tables/forecast_sum.tex
\begin{table}[ht]
\centering
\begin{tabular}{l|lrllll}
  \hline
Model & LS & p-value & maxLS & DSS & RPS & SES \\ 
  \hline
P0 & 1.71 (9) & 0.00 & 8.51 (9) & 3.24 (9) & 1.20 (8) & 17.71 (7) \\ 
  NB1 & 1.36 (8) & 0.00 & 3.85 (1) & 1.88 (6) & 1.22 (9) & 15.89 (3) \\ 
  NB2 & 1.35 (5) & 0.01 & 4.07 (8) & 1.94 (8) & 1.19 (6) & 16.23 (4) \\ 
  NB3 & 1.35 (6) & 0.01 & 4.06 (7) & 1.94 (7) & 1.19 (7) & 15.86 (2) \\ 
  ZI1 & 1.36 (7) & 0.00 & 3.86 (2) & 1.85 (3) & 1.18 (5) & 15.47 (1) \\ 
  ZI2 & 1.34 (3) & 0.02 & 4.05 (6) & 1.87 (5) & 1.13 (3) & 17.66 (6) \\ 
  ZI3 & 1.33 (1) &  & 4.04 (5) & 1.82 (1) & 1.13 (1) & 17.21 (5) \\ 
  ZI4 & 1.34 (4) & 0.03 & 3.96 (3) & 1.86 (4) & 1.14 (4) & 18.05 (8) \\ 
  ZI5 & 1.34 (2) & 0.06 & 3.97 (4) & 1.82 (2) & 1.13 (2) & 18.08 (9) \\ 
   \hline
\end{tabular}
\caption{Performance of one-step-ahead forecasts
             in terms of proper scoring rules: mean log score (LS),
             maximum log score (maxLS),
             mean Dawid-Sebastiani score (DSS),
             mean ranked probability score (RPS)
             and mean squared error score (SES).
             Ranks are shown in parantheses.
             The Monte Carlo p-values for differences in mean log scores
             are based on 9999 random permutations,
             comparing each model against the best model.} 
\label{tab:forecast}
\end{table}